# Compositional and temperature evolution of crystal structure of new thermoelectric compound LaOBiS$_{2-x}$Se$_x$

Y. Mizuguchi[1]*, A. Miura[2], A. Nishida[1], O. Miura[1], K. Tadanaga[2], N. Kumada[3], C. H. Lee[4], E. Magome[5], C. Moriyoshi[5], Y. Kuroiwa[5]

1. Department of Electrical and Electronic Engineering, Tokyo Metropolitan University, 1-1, Minami-osawa, Hachioji, Tokyo 192-0397 Japan.
2. Faculty of Engineering, Hokkaido University, Kita-13, Nishi-8, Kita-ku, Sapporo, Hokkaido 060-8628 Japan.
3. Center for Crystal Science and Technology, University of Yamanashi, 7-32 Miyamae, Kofu, Yamanashi 400-8511 Japan.
4. National Institute of Advanced Industrial Science and Technology (AIST), 1-1-1 Umezono, Tsukuba, Ibaraki 305-8568 Japan.
5. Department of Physical Science, Hiroshima University, 1-3-1 Kagamiyama, Higashihiroshima, Hiroshima 739-8526 Japan.

* Corresponding author: Yoshikazu Mizuguchi (mizugu@tmu.ac.jp)

*Abstract*

We examined the crystal structure of the new thermoelectric material LaOBiS$_{2-x}$Se$_x$, whose thermoelectric performance is enhanced by Se substitution, by using powder synchrotron X-ray diffraction and Rietveld refinement. The emergence of metallic conductivity and enhancement of the thermoelectric power factor of LaOBiS$_{2-x}$Se$_x$ can be explained with the higher in-plane chemical pressure caused by the increase of Se concentration at the in-plane Ch1 site (Ch = S, Se). High-temperature X-ray diffraction measurements for optimally substituted LaOBiSSe revealed anomalously large atomic displacement parameters ($U_{iso}$) for Bi and Ch atoms in the BiCh$_2$ conduction layers. The anisotropic analysis of the atomic displacement parameters ($U_{11}$ and $U_{33}$) for the in-plane Bi and Ch1 sites suggested that Bi atoms exhibit large atomic displacement along the *c*-axis direction above 300 K, which could be the origin of the low thermal conductivity in LaOBiSSe. The large Bi vibration along the *c*-axis direction could be related to in-plane rattling, which is a new strategy for attaining low thermal conductivity and phonon-glass-electron-crystal states.

1. Introduction

Layered bismuth chalcogenides ($BiCh_2$-based; Ch = S, Se) have been drawing considerable attention in the community of condensed matter physics and chemistry because of the superconductivity discovered in several $BiCh_2$-based compounds, such as $Bi_4O_4S_3$ and $REO_{1-x}F_xBiS_2$ (RE: rare earth).[1,2] The typical crystal structure of $BiCh_2$-based materials, which is similar to those of Cu-based or Fe-based high-$T_c$ superconductors,[3,4] is composed of alternate stacks of the $BiCh_2$ conduction layer and the electrically insulating blocking layer.[5] One of the advantages of layered structures is their great flexibility in designing new materials. The replacement of conduction and blocking layers enables to create new materials, which sometimes results in the discovery of new functionality. Consequently, layered bismuth chalcogenides are promising candidate materials to investigate novel phenomena such as possible unconventional superconductivity,[6–9] tunable Rashba-type spin splitting,[10,11] giant birefringence,[12] tunable Dirac cone,[13] or high thermoelectric performance.[14–17]

Recently, we reported promising thermoelectric properties in $LaOBiS_{2-x}Se_x$ polycrystalline samples. The dimensionless figure-of-merit ($ZT$) for $x$ = 0.8 exhibited a value of $ZT$ = 0.17 at 723 K.[15] $Z$ is calculated from the equation $Z = S^2/\rho\kappa$, where $S$, $\rho$, and $\kappa$ are the Seebeck coefficient, electrical resistivity, and thermal conductivity, respectively. Furthermore, high $ZT$ exceeding 0.36 (at ~650 K), which was recently observed in a densified (hot-pressed) sample of LaOBiSSe,[18] was achieved by enhanced metallic conductivity, large absolute Seebeck coefficient, and low thermal conductivity.

However, the origins of the LaOBiS$_{2-x}$Se$_x$ thermoelectric properties have not been clarified from the crystal structure viewpoint. In this study, we investigated the LaOBiS$_{2-x}$Se$_x$ crystal structure by using powder synchrotron X-ray diffraction and Rietveld refinement to discuss the role of Se substitution in enhancing the thermoelectric performance. Furthermore, we show the temperature evolution of the crystal structure and atomic displacement parameters of optimally substituted LaOBiSSe at high temperature to discuss the origin of the low thermal conductivity.

## 2. Experimental Details

Polycrystalline samples of LaOBiS$_{2-x}$Se$_x$ ($x$ = 0, 0.2, 0.4, 0.6, 0.8, and 1) were prepared using the solid-state-reaction method as described in Ref. 15. To investigate the crystal structure of these samples, powder synchrotron X-ray diffraction measurements were performed at room temperature (RT ~ 293 K) at the BL02B2 experimental station of SPring-8 (JASRI; Proposal No. 2014B1003 and 2014B1071). We used imaging plate of a two-dimensional detector. The interval of data points was 0.1 deg. We performed high-temperature X-ray diffraction on LaOBiSSe at RT, 300, 373, 473, 573, and 673 K (JASRI; Proposal No. 2015A1441). In the high-temperature measurements, we used a LaOBiSSe sample densified using a hot-press instrument, as we recently observed high thermoelectric performance in hot-pressed (HP) LaOBiSSe samples (denoted as HP-LaOiSSe).[18] The wavelength of the radiation beam for the measurements of as-grown LaOBiS$_{2-x}$Se$_x$ samples was 0.49542(4) Å, while that used for the high-temperature measurements of LaOBiSSe was 0.49575(8) Å. We performed Rietveld refinement (RIETAN-FP[19]) using a typical structure model for REOBiCh$_2$-type materials with a tetragonal space group of P4/nmm (see Fig. 1a).[2] The crystal structure images were drawn using VESTA.[20]

## 3. Results and discussion

### 3-1. Evolution of crystal structure in LaOBiS$_{2-x}$Se$_x$

To investigate phase purity and crystal structure of the obtained LaOBiS$_{2-x}$Se$_x$ samples, we performed synchrotron X-ray diffraction. Almost no impurity peaks were detected for $x = 0.2$–1, and the structure was refined using a single-phase analysis with the tetragonal P4/nmm model, as shown in Fig. S1. The structure for $x = 0$ was reported in our previous paper.[21] The structure parameters obtained by Rietveld analysis for LaOBiS$_{2-x}$Se$_x$ are summarized in the Supplementary Material (Table S1).

Here, we define two different chalcogen sites, Ch1 and Ch2 sites, as depicted in Fig. 1a. Fig. 1b shows the nominal $x$ dependence of the refined Se concentration at the in-plane Ch1 and out-of-plane Ch2 sites. Noticeably, Se selectively occupies the Ch1 site. In previous studies, similar site selectivity was observed in F-doped La(O,F)Bi(S,Se)$_2$.[22,23] Fig. 1c–e shows the nominal $x$ dependences of the structure parameters: (c) $a$ lattice constant, (d) $c$ lattice constant, and (e) Ch1-Bi-Ch1 angle. By increasing $x$, both the $a$ and $c$ lattice constants monotonously increase because of the lattice expansion, which can be understood by considering the difference in the ionic radius of Se$^{2-}$ (1.98 Å) and S$^{2-}$ (1.84 Å). The Ch1-Bi-Ch1 angle, which represents the flatness of the BiCh1 plane, decreases with increasing $x$, suggesting that the Se substitution results in in-plane (static) disorder. This tendency is similar to the structure evolution of the Se-substituted LaO$_{0.5}$F$_{0.5}$BiS$_{2-x}$Se$_x$ superconductors.[22] Fig. 1f–h shows three different Bi-Ch distances. The inter-plane Bi-Ch1 distance (Fig. 1f) assumes the lowest value at $x = 0.2$ and increases by increasing $x$ up to 1. The in-plane Bi-Ch1 distance (Fig. 1g) monotonously increases with increasing $x$. The Bi-Ch2 distance (Fig. 1h) decreases by increasing $x$ up to 0.6, and it increases for higher values of $x$.

In LaOBiS$_{2-x}$Se$_x$, the power factor ($S^2/\rho$) is largely enhanced by Se substitution.[15,18] When $x$ increases, $\rho$ decreases without large degradation of $S$, implying that the Se substitution does not largely affect the electron carrier concentrations, but it enhances the carrier mobility; the study on Hall measurements for LaOBiS$_{2-x}$Se$_x$ will appear in other article. To relate the evolutions of in-plane structure and physical properties, we use the concept proposed in the structurally similar La(O,F)Bi(S,Se)$_2$ superconductor family.[22] The increase in Se occupancy in the BiCh1 plane enhances the in-plane chemical pressure (CP), and metallic conductivity and bulk superconductivity are induced by the enhanced in-plane CP.[22] This can be explained by the enhancement of orbital overlap of between Bi-6p and Ch-p orbitals. The different energy level of the Se and S p orbitals can also affect the properties. We assume that the substitution of Se into Ch1 sites decreases the band gaps of LaOBiSSe, as the valence band top and the conduction band bottom are formed by Bi-Ch1 orbitals.[6,21] On the basis of this scenario, the increase in Se occupancy at the in-plane Ch1 site should be related to the emergence of metallic conductivity and the enhancement of the power factor in the present LaOBiS$_{2-x}$Se$_x$ system. However, it is quite difficult to directly discuss the obtained structure parameters and the magnitude of CP because the Bi-Ch1 distance increases with the Se concentration. Therefore, we use the equation, CP = [R$_{Bi}$ + <R$_{Ch1}$>]/[Bi-Ch1]$_{obs}$, as demonstrated in Ref. 22. In the equation, R$_{Bi}$ is the ionic radius of Bi$^{3+}$ (equal to 1.03 Å, assuming a coordination number of 6); <R$_{Ch1}$> is the average value of the ionic radius of Ch$^{2-}$ at the Ch1 site (the ionic radii of S$^{2-}$ and Se$^{2-}$ are 1.84 and 1.98 Å, respectively, assuming a coordination number of 6; in calculating <R$_{Ch1}$>, the Se occupancy at the Ch1 site, Fig. 1b, is used); [Bi-Ch1]$_{obs}$ is the in-plane Bi-Ch1 distance (Fig. 1g). The in-plane CP of LaOBiS$_{2-x}$Se$_x$, which is plotted as a function of $x$ in Fig. 2a, is clearly enhanced by increasing the value of $x$. In addition, to relate the enhanced in-plane CP to the evolution of the metallic conductivity, $\rho$ at RT for LaOBiS$_{2-x}$Se$_x$ was plotted as a function of the calculated in-plane CP (Fig. 2b).[15] The value of $\rho$(RT) largely decreases with increasing $x$ up to $x$ = 0.4, whereas low $\rho$ values are observed for $x \geq 0.4$. Therefore, the emergence of the metallic conductivity and the enhancement of the power factor may be related to the enhancement of the in-plane CP. To understand the physical properties of this system, the Se substitution appears to be more critical than the effect of the Bi-Ch1 distance elongation. Hence, tuning the in-plane CP and energy levels would be a useful strategy to design BiCh$_2$-based materials with new functionality.

**3-2. Crystal structure of LaOBiSSe at high temperatures**

Recently, we observed a high *ZT* value of 0.36 at 650 K in HP-LaOBiSSe.[18] Particularly, low thermal conductivity is essential to obtain high *ZT* values in LaOBiSSe. Here, we show the temperature evolution of the crystal structure and the atomic displacements of HP-LaOBiSSe (hot-pressed sample) at high temperatures, and discuss the relationship between low thermal conductivity and crystal structure. Fig. 3a–c shows the temperature dependences of the crystal structural parameters: (a) *a* lattice constant, (b) *c* lattice constant, and (c) Ch1-Bi-Ch1 angle. By increasing the temperature, the *a*-axis linearly expands. The *c*-axis, however, shows a non-linear expansion. The *c* lattice constant does not show a noticeable expansion below 473 K, but it expands above 473 K. The Ch1-Bi-Ch1 angle also shows a non-linear change: it increases with increasing the temperature up to 473 K, and almost saturates above this temperature. This trend indicates that the BiCh1 plane becomes flatter with increasing the temperature, although the change is not significant. Fig. 3d–f shows the temperature evolution of three different Bi-Ch distances: (d) inter-plane Bi-Ch1, (e) in-plane Bi-Ch1, and (f) Bi-Ch2 distances. The inter-plane Bi-Ch1 and in-plane Bi-Ch1 distances monotonously increase with increasing temperature. The Bi-Ch2 distance slightly decreases with the temperature up to 473 K, but it starts increasing above this value; this behavior should be related to the non-linear expansion of the *c*-axis.

In Fig. 4, the isotropic atomic displacement parameters ($U_{iso}$) for the La, Bi, Ch1, and Ch2 sites are plotted as a function of the temperature. To help the discussion of the characteristics of the atomic displacements of LaOBiSSe, we added eye-guide curves to the $U_{iso}$ temperature dependence plots. For Ch2, a kink is observed at ~473 K; note that the temperature dependence of the resistivity of LaOBiSSe also exhibits an anomaly at around this temperature.[18] This anomaly is related to the non-linear thermal expansion along the $c$-axis. We note that the $U_{iso}$ values of Bi, Ch1, and Ch2 are significantly large, compared with that observed for La. The large atomic displacements in the conduction layer are a general property of the BiCh$_2$-based family.[22,24–26] In general, $U_{iso}$ approaches zero at 0 K by extrapolating its temperature dependence. In the case of anharmonic lattice vibrations, for example with rattling atoms in a cage structure, $U_{iso}$ does not approach zero and shows a finite large $U$ value at 0 K; a similar analysis is reported in Ref. 27. For the La site, the extrapolation of the temperature dependence of $U_{iso}$ roughly approaches zero at 0 K. In contrast, the extrapolation of the temperature dependence of $U_{iso}$ for the Bi, Ch1, and Ch2 sites does not reach zero at 0 K, which indicates the possibility of unusual atomic vibrations of the Bi and Ch atoms. Particularly, the in-plane Bi and Ch1 atoms exhibit large $U_{iso}$ values. To discuss the characteristics of the atomic displacements for the Bi and Ch1 sites, we performed a Rietveld analysis with anisotropic displacement parameters ($U_{11}$ and $U_{33}$) for the in-plane Bi and Ch1 sites.

The anisotropic Rietveld analysis (Table S2) resulted in a lower $R_{wp}$ factor: for example, for the analysis of the 673 K diffraction data, the $R_{wp}$ value slightly decreased from 5.5% (isotropic analysis) to 5.4% (anisotropic analysis). The refined $U_{11}$ and $U_{33}$ for the Bi and Ch1 sites are plotted in Fig. 5a and b as functions of the temperature. In addition, to visualize the anisotropic atomic displacements at 300 and 673 K, schematic images of the crystal structure of the BiCh$_2$ conduction layer—depicted using $U_{11}$ and $U_{33}$ for the Bi and Ch1 sites and $U_{iso}$ for the Ch2 site (probability: 90%)—are displayed in Fig. 5c and d. For the Bi site, both $U_{11}$(Bi) and $U_{33}$(Bi) increase with increasing temperature, and $U_{33}$(Bi) is larger than $U_{11}$(Bi) at the entire temperature range. The temperature evolutions suggest that the thermal vibrations of Bi along the *c*-axis are enhanced at higher temperatures, and large $U_{33}$(Bi) is caused by atomic vibrations and/or Bi displacements along the *c*-axis. Atomic vibrations of Bi along the *c*-axis was more likely since the models having two Bi sites (0, 0.5, $z\pm\delta$) with half occupancy did not improve the fitting parameters of Rietveld analyses. For the Ch1 site, Below 473 K, $U_{11}$(Ch1) becomes almost constant, whereas $U_{33}$(Ch1) continues to decrease with the decrease of the temperature. Comparing Fig. 5c and d, we notice that the thermal factor for the Ch1 site changes from highly anisotropic to almost isotropic with increasing the temperature from 300 to 673 K. The anomalous evolution of $U_{11}$(Ch1) is consistent with the anomalies in the lattice constant and electrical resistivity. Therefore, the evolution of $U_{11}$ for the Ch1 site may be related to the presence of the in-plane static disorder. As a fact, the BiCh1 plane contains disorder at the Ch1 site due to the S/Se solution. Furthermore, it is possible that the disorder at Ch1 would result in a disordered local structure in the Bi-Ch1 plane. Recently, the existence of local Ch1 displacements in BiCh$_2$-based compounds have been reported using two different probes, neutron[26] and X-ray experiments[29]. Similar short-range displacements of Ch1 may exist in the present LaOBiSSe sample, but, within the present experiments and analysis, it is difficult to detect such a short-range displacement of the Ch1 site in LaOBiSSe. In that case, the Bi site may be affected to a certain extent, but the possibility of the Bi site displacements has not been proposed so far. In fact, as described above, $U_{11}$(Bi) is clearly smaller than $U_{33}$(Bi) at the entire temperature range. Therefore, we assume that the Bi site is less affected by the Ch1 site disorder, while thermal displacements of Bi are large, particularly along the *c*-axis. On the basis of the experimental results and the discussion with possible local Ch1 displacements, we propose one scenario that the Bi atoms are largely vibrating along the *c*-axis. The Bi vibration scenario is consistent with the atomic displacement parameters observed for Bi in La(O,F)BiSSe single crystals.[23]

Finally, we briefly discuss the possible mechanisms behind the low thermal conductivity in this system. In LaOBiS$_{2-x}$Se$_x$, metallic conductivity is enhanced by Se substitution, and low thermal conductivity is concurrently attained. The obtained characteristics can be ascribed to the creation of a phonon-glass-electron-crystal state, which has been considered to be a powerful strategy for designing high *ZT* materials.[29] We consider that the low thermal conductivity in LaOBiS$_{2-x}$Se$_x$ is related to the large atomic displacement parameters in the Bi-Ch1 plane. Furthermore, on the basis of the anisotropic analysis of the atomic displacement parameters for LaOBiSSe, we assume that the Bi and Ch1 atoms anomalously vibrate along the *c*-axis direction, as rattling atoms in a cage structure (for example, in skutterudite compounds[30]). The crystal structure of LaOBiS$_{2-x}$Se$_x$, however, is not a cage structure; hence, the observed rattling-like vibration occurs within a plane (two-dimensional) structure. Recently, a high *ZT* value of 0.7 was reported in tetrahedrite compounds (Cu$_{10}$Tr$_2$Sb$_4$S$_{13}$; Tr denotes the transition metal) with low thermal conductivity.[31,32] In the tetrahedrite family, the low thermal conductivity is achieved by lattice anharmonicity of the Cu atoms: in other words, *in-plane rattling* of Cu atoms at Cu-S planes is the origin of the low thermal conductivity.[31,32] We assume that the low thermal conductivity in LaOBiSSe would also result from a mechanism similar to that observed in the tetrahedrite family. If the in-plane rattling can generally suppress the thermal conductivity of compounds with a plane structure, we obtain a new strategy for designing high-performance thermoelectric materials with a phonon-glass-electron-crystal state. To elucidate the assumption above, neutron measurements or other measurements more sensitive to phonons (lattice anharmonicity) are needed.

## 4. Conclusions

We examined the crystal structure of a new thermoelectric material family, LaOBiS$_{2-x}$Se$_x$, using powder synchrotron X-ray diffraction. Upon substitution, Se selectively occupies the in-plane Ch1 site. Although the Se substitution induces buckling of the Bi-Ch1 plane, its intensification results in the enhancement of the metallic conductivity and thermoelectric power factor. We explained that the enhancement of the power factor was attained owing to the increased in-plane chemical pressure. To investigate the origin of the low thermal conductivity of LaOBiSSe, we analysed the high-temperature crystal structure and atomic displacement parameters. LaOBiSSe exhibits anisotropic thermal expansion. Below 473 K, only the *a*-axis expands, while both *a*-axis and *c*-axis expand above 473 K. We revealed anomalously large atomic displacement parameters ($U_{iso}$) for the Bi and Ch atoms in the Bi-Ch1 plane. The temperature dependences of $U_{iso}$ suggest that the temperature evolutions of the atomic vibrations of the Bi-Ch1 plane in the BiCh$_2$ layers are anomalous and similar to that of the rattling atoms in a cage structure. The anisotropic analysis of the atomic displacement parameters ($U_{11}$ and $U_{33}$) for the in-plane Bi and Ch1 sites suggests that Bi atoms exhibit large atomic displacement along the *c*-axis direction, which should presumably be the origin of the low thermal conductivity in LaOBiSSe. In addition, we assume that the large Bi and Ch1 vibrations are related to the in-plane rattling, which is a new strategy for designing low-$\kappa$ thermoelectric materials, as observed in Cu atoms of tetrahedrite compounds.

### Acknowledgements

This work was partly supported by Grant-in-Aid for Young Scientist (A) (25707031), Grant-in-Aid for challenging Exploratory Research (26600077 and 15K14113), and Grant-in-Aid for Scientific Research on Innovative Areas

**References**

1  Y. Mizuguchi, H. Fujihisa, Y. Gotoh, K. Suzuki, H. Usui, K. Kuroki, S. Demura, Y. Takano, H. Izawa, and O. Miura, *Phys. Rev. B* 86, 220510 (2012).

2  Y. Mizuguchi, S. Demura, K. Deguchi, Y. Takano, H. Fujihisa, Y. Gotoh, H. Izawa, and O. Miura, *J. Phys. Soc. Jpn.* 81, 114725 (2012).

3  J. B. Bednorz and K. Müller, *Z. Physik B Condens. Matter* 64, 189 (1986).

4  Y. Kamihara, T. Watanabe, M. Hirano, and H. Hosono, *J. Am. Chem. Soc.* 130, 3296 (2008).

5  Y. Mizuguchi, *J. Phys. Chem. Solid* 84, 34 (2015).

6  H. Usui, K. Suzuki, K. Kuroki, *Phys. Rev. B* 86, 220501 (2012).

7  Y. Gao, T. Zhou, H. Huang, P. Tong, and Q. H. Wang, *Phys. Rev. B* 90, 054518 (2014).

8  J. Liu, D. Fang, Z. Wang, J. Xing, Z. Du, X. Zhu, H. Yang, and H. H. Wen, *EPL* 106, 67002 (2014).

9  Y. Yang, W. S. Wang, Y. Y. Xiang, Z. Z. Li, and Q. H. Wang, *Phys. Rev. B* 88, 094519 (2013).

10 Q. Liu, Y. Guo, and A. J. Freeman, *Nano Lett.* 13, 5264 (2013).

11 Y. Ma, Y. Dai, N. Yin, T. Jing, and B. Huang, *J. Mater. Chem. C*, 2014, 2, 8539.

12 H. Wang, *Chin. Phys. Lett.* 31, 047802 (2014).

13 X. Y. Dong, J. F. Wang, R. X. Zhang, W. H. Duan, B. F. Zhu, J. Sofo, and C. X. Liu, arXiv:1409.3641.

14 A. Omachi, J. Kajitani, T. Hiroi, O. Miura, and Y. Mizuguchi, *J. Appl. Phys.* 115, 083909 (2014).

15 Y. Mizuguchi, A. Omachi, Y. Goto, Y. Kamihara, M. Matoba, T. Hiroi, J. Kajitani, and O. Miura, *J. Appl. Phys.* 116, 163915 (2014).

16 Y. L. Sun, A. Ablimit, H. F. Zhai, J. K. Bao, Z. T. Tang, X. B. Wang, N. L. Wang, C. M. Feng, and G. H. Cao, *Inorg. Chem.* 53, 11125 (2014).

17 Y. Goto, J. Kajitani, Y. Mizuguchi, Y. Kamihara, and M. Matoba, *J. Phys. Soc. Jpn.* 84, 085003 (2015).

18 A. Nishida, O. Miura, C. H. Lee, and Y. Mizuguchi, *Appl. Phys. Express* 8, 111801 (2015).

19 F. Izumi, and K. Momma, *Solid State Phenom.* 130, 15 (2007).

20 K. Momma, and F. Izumi, *J. Appl. Crystallogr.* 41, 653 (2008).

21 A. Miura, Y. Mizuguchi, T. Takei, N. Kumada, E. Magome, C. Moriyoshi, Y. Kuroiwa, and K. Tadanaga, Solid State Commun. 227, 19 (2016).

22 Y. Mizuguchi, A. Miura, J. Kajitani, T. Hiroi, O. Miura, K. Tadanaga, N. Kumada, E. Magome, C. Moriyoshi, and Y. Kuroiwa, *Sci. Rep.* 5, 14968 (2015).

23 M. Tanaka, T. Yamaki, Y. Matsushita, M. Fujioka, S. J. Denholme, T. Yamaguchi, H. Takeya, and Y. Takano, *Appl. Phys. Lett.* 106, 112601 (2015).

24 E. Paris, B. Joseph, A. Iadecola, T. Sugimoto, L. Olivi, S. Demura, Y. Mizuguchi, Y. Takano, T. Mizokawa, and N. L. Saini, *J. Phys.: Condens. Matter* 26, 435701 (2014).

25 Y. Mizuguchi, E. Paris, T. Sugimoto, A. Iadecola, J. Kajitani, O. Miura, T. Mizokawa, and N. L. Saini, *Phys. Chem. Chem. Phys.* 17, 22090 (2015).

26 A. Athauda, J. Yang, S. Lee, Y. Mizuguchi, K. Deguchi, Y. Takano, O. Miura, and D. Louca, *Phys. Rev. B* 91, 144112 (2014).

27 Y. Akizuki, I. Yamada, K. Fujita, K. Taga, T. Kawakami, M. Mizumaki, and K. Tanaka, *Angew. Chem. Int. Ed.* 54, 1 (2015).

28 R. Sagayama, H. Sagayama, R. Kumai, Y. Murakami, T. Asano, J. Kajitani, R. Higashinaka, T. D. Matsuda, and Y. Aoki, *J. Phys. Soc. Jpn.* 84, 123703 (2015).

29 G. A. Slack, *CRC Handbook of Thermoelectrics, ed. DM Rowe*, pp. 407 (1995).

30 B. C. Sales, D. Mandrus, and R. K. Williams, *Science* 272, 1325 (1996).

31 K. Suekuni, K. Tsuruta, T. Ariga, and M. Koyano, *Appl. Phys. Express* 5, 051201 (2012).

32 K. Suekuni, K. Tsuruta, M. Kunii, H. Nishiate, E. Nishibori, S. Maki, M. Ohta, A. Yamamoto, and M. Koyano, *J. Appl. Phys.* 113, 043712 (2013).

33 See supplemental material at [URL will be inserted by AIP] for a typical X-ray diffraction pattern and crystal structural parameters. In Fig. S1, we show the powder X-ray diffraction pattern and the fitting result of Rietveld analysis for LaOBiSSe (x = 1). In table S1 and S2, we show the refined structural parameters for LaOBiS$_{2-x}$Se$_x$ at room temperature and for LaOBiSSe at high temperatures, respectively.

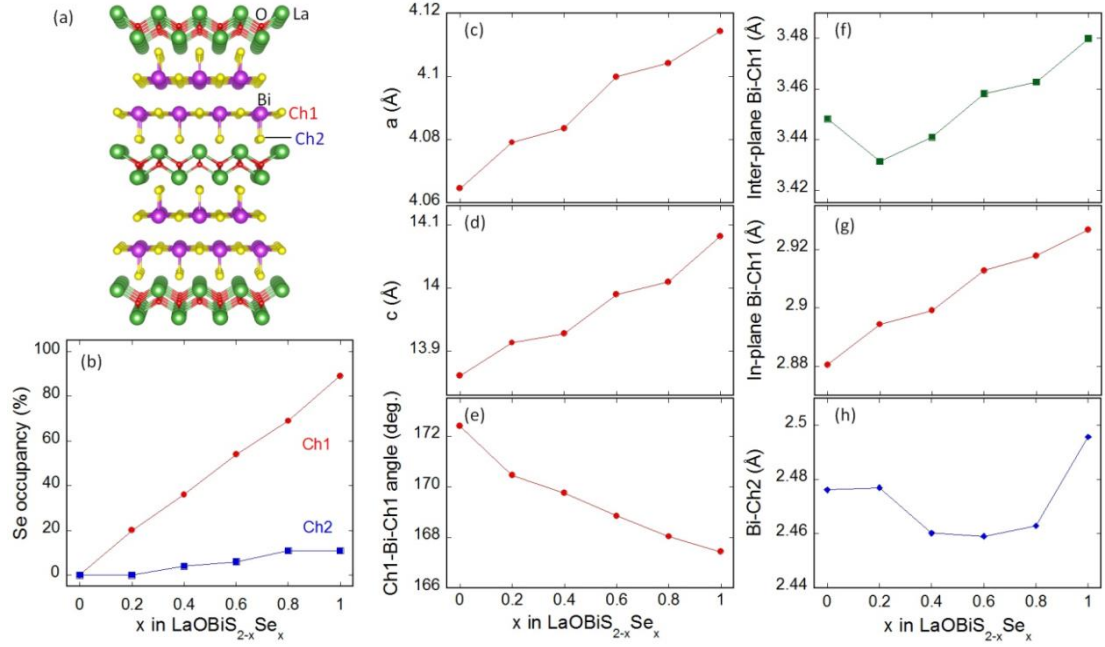

Fig. 1. (a) Schematic of crystal structure of LaOBiCh$_2$ (Ch: S, Se). (b) Refined Se concentration at the Ch1 and Ch2 sites of LaOBiS$_{2-x}$Se$_x$ as a function of nominal Se concentration ($x$). (c–h) Nominal $x$ dependence of refined crystal structural parameters for LaOBiS$_{2-x}$Se$_x$: (c) $a$ lattice constant, (d) $c$ lattice constant, (e) Ch1-Bi-Ch1 angle, (f) inter-plane Bi-Ch1 distance, (g) in-plane BiCh1 distance, and (h) Bi-Ch2 distances.

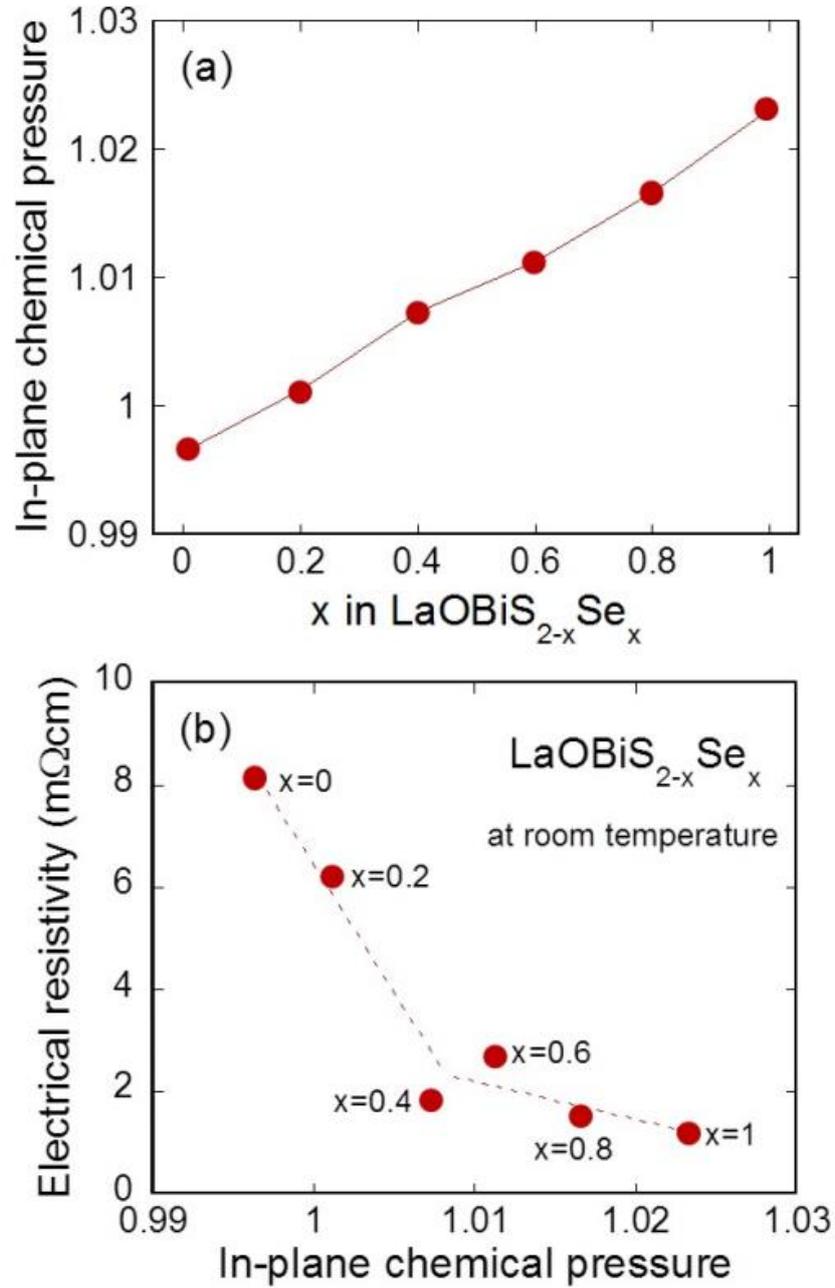

Fig. 2. (a) Se concentration dependence of in-plane chemical pressure (dimensionless) in LaOBiS$_{2-x}$Se$_x$. (b) In-plane chemical pressure dependence of electrical resistivity ($\rho$) at room temperature for in LaOBiS$_{2-x}$Se$_x$.[15] The dashed lines are eye-guides.

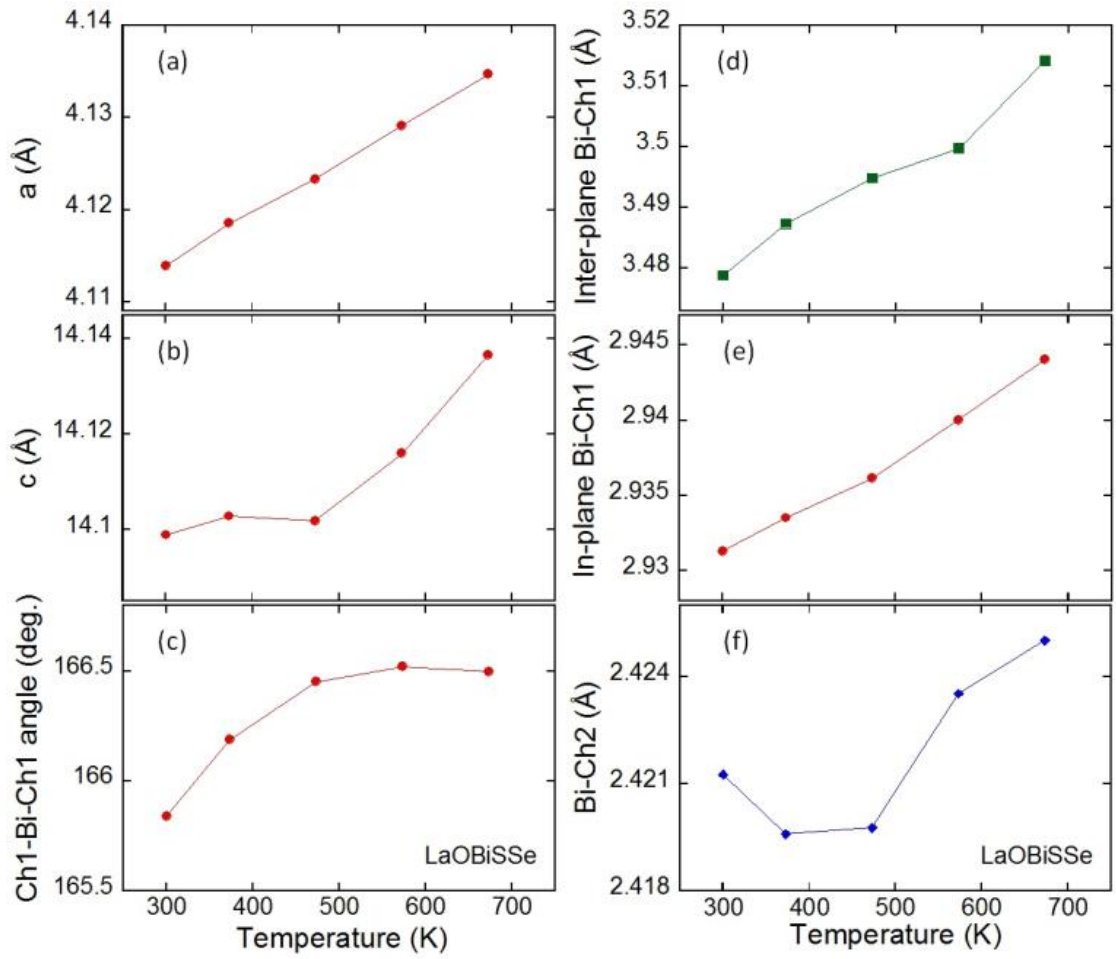

Fig. 3. Temperature dependences of structure parameters for LaOBiSSe: (a) *a* lattice constant, (b) *c* lattice constant, (c) Ch1-Bi-Ch1 angle, (d) inter-plane Bi-Ch1 distance, (e) in-plane BiCh1 distance, and (f) Bi-Ch2 distances.

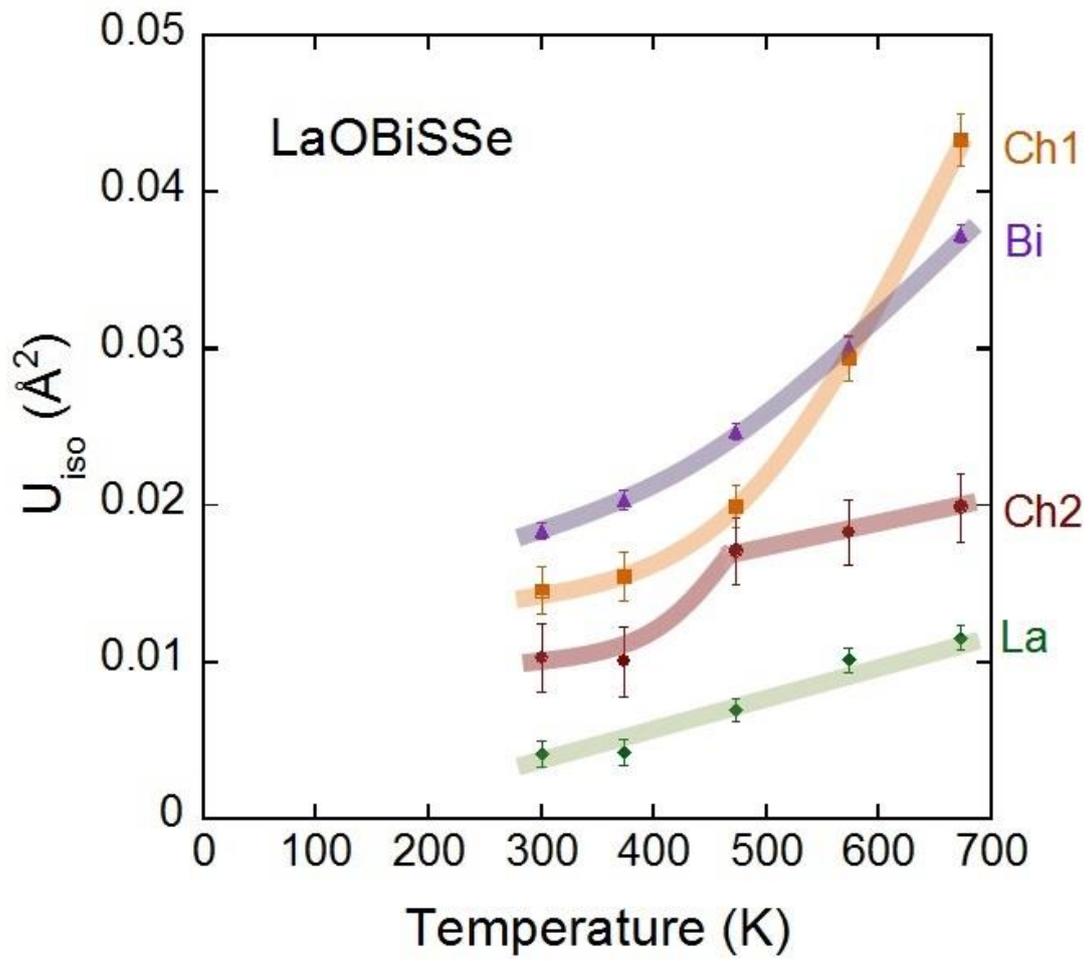

Fig. 4. Temperature dependences of isotropic atomic displacement parameters ($U_{iso}$) for the La, Bi, Ch1, and Ch2 sites of LaOBiSSe.

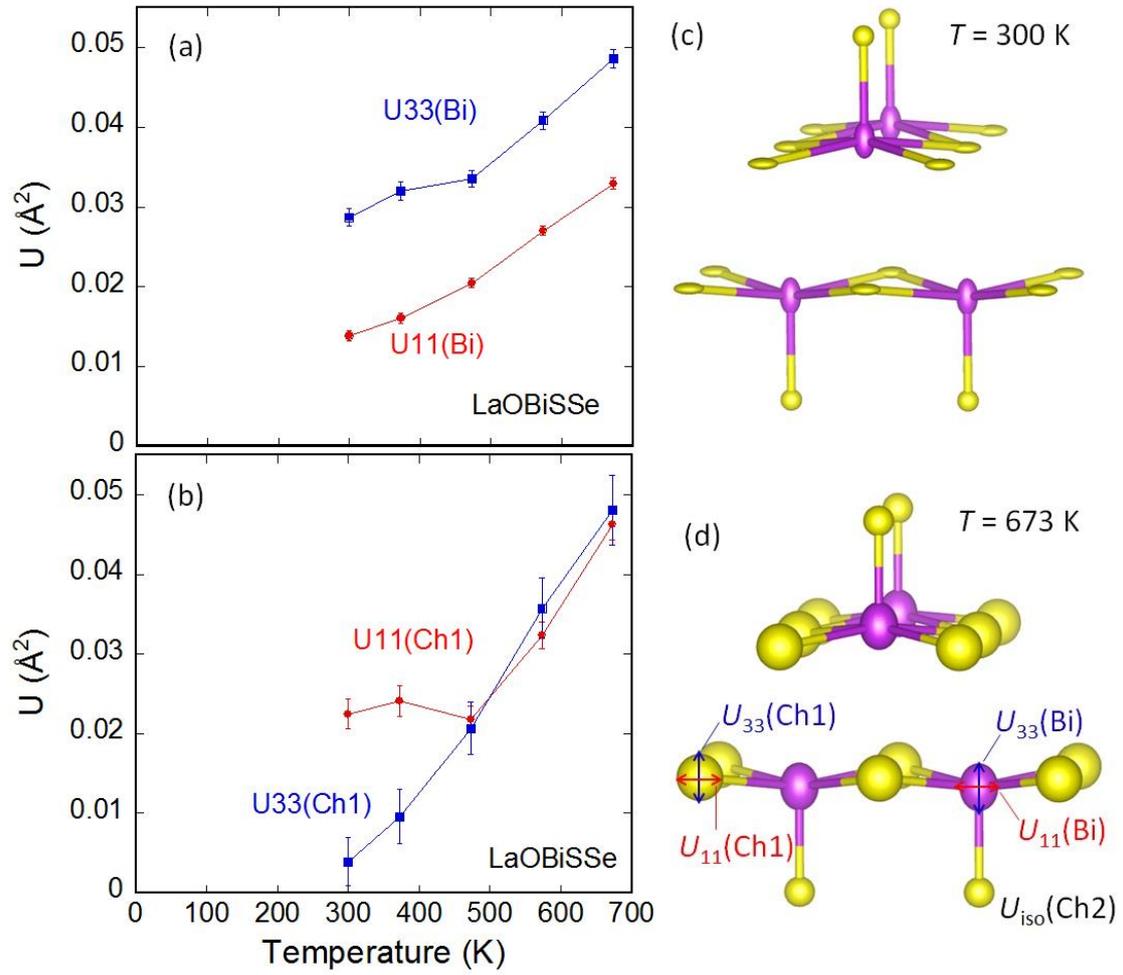

Fig. 5. Evolution of in-plane atomic displacement parameters ($U_{11}$ and $U_{33}$) for (a) Bi and (b) Ch1 sites of LaOBiSSe, examined by using anisotropic analysis of atomic displacement parameters. (c,d) Schematic images of the crystal structure of the BiCh$_2$ conduction layer depicted using $U_{11}$ and $U_{33}$ at (c) 300 K and (d) 673 K for the Bi and Ch1 sites (probability: 90%). For the Ch2 site, $U_{iso}$ was used.

# Supplemental Materials

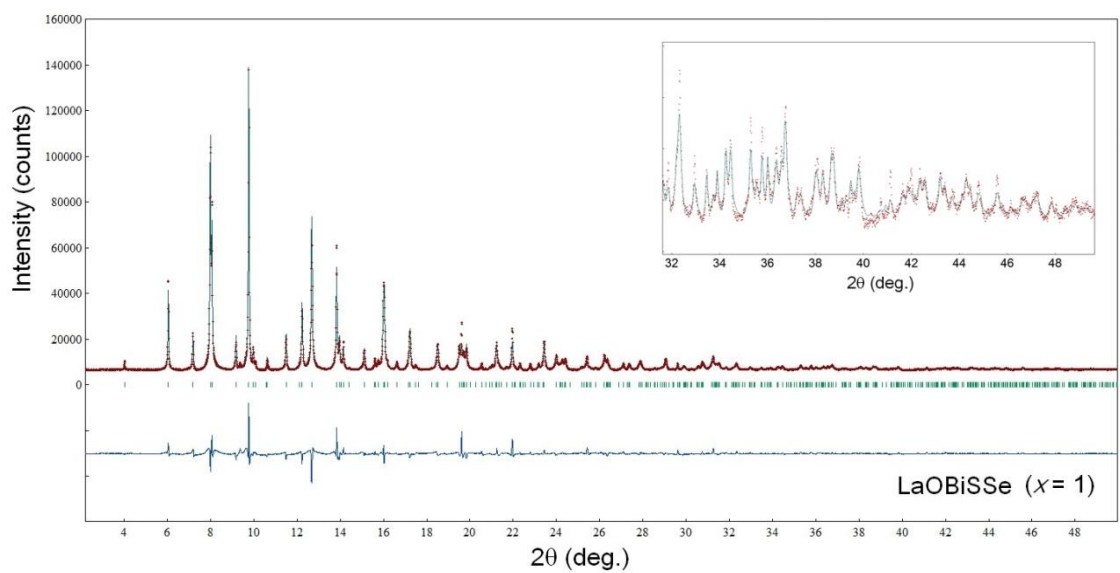

Fig. S1. Powder X-ray diffraction pattern and the fitting result of Rietveld analysis for LaOBiSSe ($x$ = 1). The inset shows enlarged pattern at higher $2\theta$ angles.

Table S1. Crystal structure parameters of LaOBiS$_{2-x}$Se$_x$ obtained from Rietveld refinements.

| $x$ | $T$ (K) | $a$ (Å) / $c$ (Å) | $z$ (La) / $U_{iso}$ (Å$^2$) | $z$ (Bi) / $U_{iso}$ (Å$^2$) | $z$ (Ch1) / $U_{iso}$ (Å$^2$) | $z$ (Ch2) / $U_{iso}$ (Å$^2$) | g_Se1 (%) | $R_{wp}$ (%) |
|---|---|---|---|---|---|---|---|---|
| 0 | R.T. | 4.06456(3) / 13.8614(2) | 0.0899(1) / 0.0054(4) | 0.63130(1) / 0.0128(3) | 0.3825(4) / 0.029(2) | 0.8099(3) / 0.007(1) | 0 | 5.4 |
| 0.2 | R.T. | 4.07913(6) / 13.9134(2) | 0.08952(8) / 0.0052(5) | 0.63197(9) / 0.0169(3) | 0.3853(3) / 0.027(2) | 0.8100(4) / 0.008(2) | 20(1) | 3.8 |
| 0.4 | R.T. | 4.08357(7) / 13.9278(3) | 0.0886(1) / 0.0038(6) | 0.6328(1) / 0.0165(4) | 0.3858(3) / 0.021(2) | 0.8095(5) / 0.018(2) | 36(1) | 6.0 |
| 0.6 | R.T. | 4.09987(7) / 13.9898(3) | 0.0879(1) / 0.0039(7) | 0.6337(1) / 0.0170(5) | 0.3865(3) / 0.020(2) | 0.8095(6) / 0.019(3) | 54(1) | 7.4 |
| 0.8 | R.T. | 4.10420(8) / 14.0097(4) | 0.0875(1) / 0.0036(8) | 0.6344(1) / 0.0162(5) | 0.3873(3) / 0.015(2) | 0.8102(6) / 0.024(3) | 69(1) | 7.0 |
| 1 | R.T. | 4.11431(6) / 14.0818(3) | 0.0876(1) / 0.0065(8) | 0.6349(1) / 0.0169(5) | 0.3878(3) / 0.023(2) | 0.8122(5) / 0.008(2) | 89(1) | 5.2 |
| 1 (HP) | 300 | 4.11382(9) / 14.0989(4) | 0.0860(1) / 0.0041(8) | 0.6362(1) / 0.0184(6) | 0.3894(3) / 0.015(2) | 0.8079(5) / 0.010(2) | 89(1) | 6.6 |
| 1 (HP) | 373 | 4.11854(9) / 14.1028(4) | 0.0860(1) / 0.0043(9) | 0.6362(1) / 0.0204(6) | 0.3889(3) / 0.016(2) | 0.8077(5) / 0.010(2) | 89(1) (fixed) | 6.5 |
| 1 (HP) | 473 | 4.12331(7) / 14.1018(3) | 0.0856(1) / 0.0070(8) | 0.6362(1) / 0.0248(5) | 0.3884(2) / 0.020(1) | 0.8078(4) / 0.017(2) | 89(1) (fixed) | 5.7 |
| 1 (HP) | 573 | 4.12908(6) / 14.1160(3) | 0.0860(1) / 0.0101(7) | 0.6362(1) / 0.0302(5) | 0.3883(2) / 0.029(1) | 0.8079(4) / 0.018(2) | 89(1) (fixed) | 5.3 |
| 1 (HP) | 673 | 4.13469(7) / 14.1367(3) | 0.0860(1) / 0.0115(8) | 0.6365(1) / 0.0374(6) | 0.3880(2) / 0.043(2) | 0.8081(4) / 0.020(2) | 89(1) (fixed) | 5.5 |

Table S2. Crystal structure parameters of LaOBiSSe at high temperatures obtained from Rietveld refinements with anisotropic displacement parameters for Bi and Ch1 sites.

| $T$ (K) | $a$ (Å)<br>$c$ (Å) | $z$ (La)<br>$U_{iso}$ (Å$^2$) | $z$ (Bi)<br>$U_{11}$ (Å$^2$)<br>$U_{33}$ (Å$^2$) | $z$ (Ch1)<br>$U_{11}$ (Å$^2$)<br>$U_{33}$ (Å$^2$) | $z$ (Ch2)<br>$U_{iso}$ (Å$^2$) | $R_{wp}$ (%) |
|---|---|---|---|---|---|---|
| 300 | 4.11389(9)<br>14.0989(4) | 0.0858(1)<br>0.0037(8) | 0.6361(1)<br>0.0138(6)<br>0.029(1) | 0.3896(2)<br>0.023(2)<br>0.004(3) | 0.8071(5)<br>0.008(2) | 6.5 |
| 373 | 4.11860(8)<br>14.1031(4) | 0.0859(1)<br>0.0046(9) | 0.6362(1)<br>0.0160(7)<br>0.032(1) | 0.3888(3)<br>0.024(2)<br>0.010(3) | 0.8072(5)<br>0.008(2) | 6.4 |
| 473 | 4.12343(7)<br>14.1020(3) | 0.0860(1)<br>0.0066(7) | 0.6363(1)<br>0.0205(6)<br>0.034(1) | 0.3881(2)<br>0.022(2)<br>0.021(3) | 0.8081(4)<br>0.016(2) | 5.6 |
| 573 | 4.12918(6)<br>14.1165(3) | 0.0860(1)<br>0.0107(7) | 0.6363(1)<br>0.0270(6)<br>0.041(1) | 0.3880(2)<br>0.032(2)<br>0.036(4) | 0.8079(4)<br>0.019(2) | 5.2 |
| 673 | 4.13467(6)<br>14.1367(3) | 0.0859(1)<br>0.0111(8) | 0.6366(1)<br>0.0330(7)<br>0.049(1) | 0.3877(3)<br>0.046(2)<br>0.048(4) | 0.8079(4)<br>0.020(2) | 5.4 |